\documentclass[a4paper,11pt]{article}

\usepackage{jheppub}

\usepackage{amsmath}
\usepackage{bm}
\usepackage{dcolumn}
\usepackage{graphicx}
\usepackage[normalem]{ulem}

\newcommand{\tr}{\ensuremath{\mathrm{Tr}}}

\begin{document}

\title{Axion phenomenology and $\theta$-dependence 
from $N_f = 2+1$ lattice QCD}

\author[a,b]{Claudio Bonati,}
\emailAdd{claudio.bonati@df.unipi.it}
\affiliation[a]{Dipartimento di Fisica dell'Universit\`a di Pisa, Largo Pontecorvo 3, I-56127 Pisa, Italy}
\affiliation[b]{INFN sezione di Pisa, Largo Pontecorvo 3, I-56127 Pisa, Italy}

\author[a,b]{Massimo D'Elia,}
\emailAdd{massimo.delia@unipi.it}

\author[a,b]{Marco Mariti,}
\emailAdd{mariti@df.unipi.it}

\author[c,d]{Guido Martinelli,}
\emailAdd{guido.martinelli@roma1.infn.it}
\affiliation[c]{Dipartimento di Fisica dell'Universit\`a di Roma ``La Sapienza'', 
Piazzale Aldo Moro 5, I-00185 Roma, Italy}
\affiliation[d]{INFN Sezione di Roma La Sapienza, Piazzale Aldo Moro 5, I-00185 Roma, Italy}

\author[a,b]{Michele Mesiti,}
\emailAdd{mesiti@pi.infn.it}

\author[b]{Francesco Negro,}
\emailAdd{fnegro@pi.infn.it}

\author[e]{Francesco Sanfilippo,}
\emailAdd{f.sanfilippo@soton.ac.uk}
\affiliation[e]{School of Physics and Astronomy, University of Southampton, SO17 1BJ Southampton, United Kingdom}

\author[f]{Giovanni Villadoro}
\emailAdd{giovanni.villadoro@ictp.it}
\affiliation[f]{Abdus Salam International Centre for Theoretical Physics, Strada Costiera 11, 34151, Trieste, Italy}

\date{\today}

\abstract{We investigate the topological properties of $N_f = 2+1$ QCD with physical
quark masses, both at zero and finite temperature. We adopt stout improved
staggered fermions and explore a range of lattice spacings $a \sim 0.05 -
0.12$ fm.  At zero temperature we estimate both finite size and finite cut-off
effects, comparing our continuum extrapolated results for the topological
susceptibility $\chi$ with predictions from chiral perturbation theory. At
finite temperature, we explore a region going from $T_c$ up to around 
$4\, T_c$, where we provide continuum extrapolated results for the topological
susceptibility and for the fourth moment of the topological charge
distribution. While the latter converges to the dilute instanton gas prediction
the former differs strongly both in the size and in the temperature dependence.
This results in a shift of the axion dark matter window
of almost one order of magnitude with respect to the instanton computation.}


\maketitle
\flushbottom

\section{Introduction}\label{intro}

Axions are among the most interesting candidates for physics beyond the
Standard Model. Their existence has been advocated long
ago~\cite{Peccei:1977hh, Peccei:1977ur, Wilczek:1977pj, Weinberg:1977ma} as a
solution to the so-called strong-CP problem through the Peccei-Quinn
(PQ) mechanism.  It was soon realized that they could also explain the
observed dark matter abundance of the visible Universe~\cite{Preskill:1982cy,
Abbott:1982af, Dine:1982ah}.  However, a reliable computation of the axion
relic density requires a quantitative estimate of the parameters entering the
effective potential of the axion field, in particular its mass and
self-couplings as a function of the temperature $T$ of the thermal bath.

The purpose of this study is to obtain predictions from the numerical
simulations of Quantum Chromodynamics (QCD) on a lattice.  Our
results, which are summarized at the end of this section, suggest a
possible shift of the axion dark matter window by almost one order of
magnitude with respect to instanton computations.  This shift is a
consequence of the much slower decrease of the axion mass with the
temperature in comparison to the dilute instanton gas prediction.  Our
present simulations are however limited to a range of temperatures not
exceeding 600~MeV: the main obstruction is represented by the freezing
of the topological modes on fine lattices, which afflicts present
lattice QCD algorithms. For a more complete understanding of axion
dynamics at finite $T$, in the future a major effort must be
undertaken to reach higher temperatures.

\subsection{General framework}

Given the strong bounds on its couplings, the axion field can be
safely treated as a non-dynamical external field.  Its potential is
completely determined by the dependence of the QCD partition function
on the $\theta$-angle, which enters the pure gauge part of the QCD
Euclidean Lagrangian as
\begin{equation}\label{lagrangian}
\mathcal{L}_\theta = \frac{1}{4} F_{\mu\nu}^a(x)F_{\mu\nu}^a(x) - i
\theta q(x),
\end{equation}
where 
\begin{equation}\label{topchden}
q(x)=\frac{g^2}{64\pi^2} 
\epsilon_{\mu\nu\rho\sigma} F_{\mu\nu}^a(x) F_{\rho\sigma}^a(x)
\end{equation}
is the topological charge density. The $\theta$-dependent part of the
free energy density can be parametrized as follows
\begin{equation}
\mathcal{F}(\theta,T)\equiv F(\theta,T)-F(0,T)={1\over 2} \chi(T)
\theta^2 s(\theta,T)
\end{equation}
where $\chi(T)$ is the topological susceptibility at $\theta=0$,
\begin{equation}
\chi = \int d^4 x \langle q(x)q(0) \rangle_{\theta=0} 
= {\langle Q^2 \rangle_{\theta=0} \over {\cal V}} \label{chidef}
\end{equation}
($Q = \int d^4 x\, q(x)$ is the global topological charge and
$\mathcal{V}=V/T$), while $s(\theta,T)$ is a dimensionless even
function of $\theta$ such that $s(0,T)=1$. The quadratic term in
$\theta$, $\chi(T)$, is proportional to the axion mass, while
non-linear corrections in $\theta^2$, contained in $s(\theta,T)$,
provide information about axion interactions.  In particular, assuming
analyticity around $\theta=0$, $s(\theta,T)$ can be expanded as
follows \cite{Vicari:2008jw}
\begin{equation}\label{stheta}
s(\theta,T) = 1 + b_2(T) \theta^2 + b_4(T) \theta^4 + \cdots,
\end{equation}
where the coefficients $b_n$ are proportional to the cumulants of the
topological charge distribution. For instance $b_2$, which is related
to quartic interactions terms in the axion potential, can be expressed
as
\begin{equation}\label{eq:b2}
b_2=-\frac{\langle Q^4\rangle_{\theta=0}-
         3\langle Q^2\rangle^2_{\theta=0}}{12\langle Q^2\rangle_{\theta=0}} \, .
\end{equation}

The function ${\cal F}(\theta,T)$, related to the topological
properties of QCD, is of non-perturbative nature and hence not easy to
predict reliably with analytic methods. This is possible only in some
specific regimes: chiral perturbation theory (ChPT) represents a valid
approach only in the low temperature phase; at high-$T$, instead, a
possible analytic approach is the Dilute Instanton Gas Approximation
(DIGA).  DIGA predictions can in fact be classified in two groups:
those that make only use of the DIGA hypothesis itself (i.e. that just
relies on the existence of weakly interacting objects of topological
charge one), and those that exploit also perturbation theory, the
latter being expected to hold only at asymptotically high values of
$T$.  Using only the dilute gas approximation one can show that the
$\theta$-dependence of the free energy is of the form (see
e.g. \cite{Gross:1980br,Schafer:1996wv})
\begin{equation}\label{eq:inst_gas}
F(\theta,T)-F(0,T)\simeq \chi(T)(1-\cos\theta)\ ,
\end{equation} 
and thus $b_2^{DIGA}=-1/12$, $b_4^{DIGA}=1/360$ and so on. Using also
perturbation theory it is possible to obtain an explicit form for the
dependence of the topological susceptibility on the temperature. To
leading order, for $N_f^{(l)}$ light quark flavors of mass $m_l$, one
obtains (see e.g. \cite{Gross:1980br,Schafer:1996wv})
\begin{equation}\label{eq:chi_inst_pert}
\chi(T)\sim T^{4-\beta_0}\left(\frac{m_l}{T}\right)^{N_f^{(l)}}\ ,
\end{equation}
where $\beta_0=11N_c/3-2N_f/3$. Only part of the NLO corrections to
this expression are known (see \cite{Morris:1984zi} or
\cite{Ringwald:1999ze} for a summary, \cite{Borsanyi:2015cka} for the
$N_f=0$ case).

As an alternative, a fully non-perturbative approach, which is based
completely on the first principles of the theory, is represented by
lattice QCD simulations.  In fact, extensive studies have been carried
out regarding the $\theta$-dependence of pure gauge theories. It was
shown in Ref.~\cite{Bonati:2013tt}, and later confirmed in
Refs.~\cite{Bonati:2015uga, Borsanyi:2015cka, Xiong:2015dya}, that the
form of the free energy in Eq.~\eqref{eq:inst_gas} describes with high
precision the physics of the system for $T\gtrsim 1.15~T_c$, while for
$T\lesssim T_c$ everything is basically independent of the
temperature, thus strengthening the conclusion $\chi(T<T_c)\approx
\chi(T=0)$ obtained in previous studies \cite{Alles:1996nm,
  Alles:2000cg,Gattringer:2002mr, Lucini:2004yh, DelDebbio:2004rw}.
In Refs.~\cite{Berkowitz:2015aua, Borsanyi:2015cka} it was also shown
that the temperature dependence of the topological susceptibility is
correctly reproduced by Eq.~\eqref{eq:chi_inst_pert} for temperatures
just above $T_c$, even if the overall normalization is about a factor
ten larger than the perturbative prediction.

A realistic study of $\theta$-dependence aimed at being relevant to
axion phenomenology requires the numerical simulation of lattice QCD
including dynamical quarks with physical masses.  Apart from the usual
computational burden involved in the numerical simulation of light
quarks, that represents a challenge from at least two different but
interrelated points of view.

Because of the strict connection, in the presence of light fermions,
between the topological content of gauge configurations and the
spectrum of the fermion matrix (in particular regarding the presence
of zero modes), a reliable study of topological quantities requires a
discretization of the theory in which the chiral properties of
fermions fields are correctly implemented.  For standard
discretizations, such properties are recovered only for small enough
lattice spacings, so that a careful investigation of the continuum
limit becomes essential.  Indeed only recently it was possible to
measure the dependence of the topological susceptibility on the quark
masses to a sufficient accuracy to be compared with the prediction of
chiral perturbation theory~\cite{Bazavov:2010xr, Bazavov:2012xda,
  Cichy:2013rra, Bruno:2014ova, Fukaya:2014zda}.

On the other hand, it is well known that, as the continuum limit is
approached, it becomes increasingly difficult to correctly sample the
topological charge distribution, because of the large energy barriers
developing between configurations belonging to different homotopy
classes, i.e.~to different topological sectors, which can be hardly
crossed by standard algorithms~\cite{Alles:1996vn, DelDebbio:2002xa,
DelDebbio:2004xh, Schaefer:2010hu, Kitano:2015fla}. That causes a loss of ergodicity
which, in principle, can spoil any effort to approach the continuum
limit itself.

Combining these two problems together, the fact that a proper window
exists, in which the continuum limit can be taken and still
topological modes can be correctly sampled by current state-of-the-art
algorithms, is highly non-trivial.  In a finite temperature study,
since the equilibrium temperature is related to the inverse temporal
extent of the lattice by $T = 1/(N_t a)$, the fact that one cannot
explore arbitrarily small values of the lattice spacing $a$, because
of the above mentioned sampling problem, limits the range of
explorable temperatures from above.

In the present study we show that, making use of current
state-of-the-art algorithms, one can obtain continuum extrapolated
results for the $\theta$-dependence of QCD at the physical point, in a
range of temperatures going up to about 4 $T_c$, where $T_c \sim 155$
MeV is the pseudo-critical temperature at which chiral symmetry
restoration takes place.  Then we discuss the consequences of such
results to axion phenomenology in a cosmological context.

Our investigation is based on the numerical simulation of $N_f = 2+1$
QCD, adopting a stout staggered fermions discretization with physical
values of the quark masses and the tree level improved Symanzik gauge
action.  First, we consider simulations at zero temperature and
various different values of the lattice spacing, in a range $\sim 0.05
- 0.12$ fm and staying on a line of constant physics, in order to
identify a proper scaling window where the continuum limit can be
taken without incurring in severe problems with the freezing of
topological modes. Results are then successfully compared to the
predictions of chiral perturbation theory. We also show that, for
lattice spacings smaller than those explored by us, the freezing
problem becomes severe, making the standard Rational Hybrid
Monte-Carlo (RHMC) algorithm useless.

For a restricted set of lattice spacings, belonging to the scaling
window mentioned above, we perform finite temperature simulations,
obtaining continuum extrapolated results for $\chi$ and $b_2$ in a
range of $T$ going up to around 600 MeV.  These results are then taken
as an input to fix the parameters of the axion potential in the same
range of temperatures and perform a phenomenological analysis.

\subsection{Summary of main results and paper organization}

Our main results are the following. Up to $T_c$ the topological
susceptibility is almost constant, and compatible with the prediction
from ChPT. Above $T_c$ the value of $b_2$ rapidly converges to what
predicted by DIGA computations; on the contrary the dependence of the
topological susceptibility on $T$ shows significant deviations from
DIGA.  This has a significant impact on axion phenomenology, in
particular it results in a shift of the axion dark matter window by
almost one order of magnitude with respect to the instanton
computation.  Since the $T$-dependence is much milder than expected
from instanton calculations, it becomes crucial, for future studies,
to investigate the system for higher values of $T$, something which
also claims for the inclusion of dynamical charm quarks and for
improved algorithms, capable to defeat or at least alleviate the
problem of the freezing of topological modes.

The paper is organized as follows.  In Section~\ref{sec:setup}, we
discuss the setup of our numerical simulations, in particular the
lattice discretization adopted and the technique used to extract the
topological content of gauge configurations.  In
Section~\ref{sec:results} we present our numerical analysis and
continuum extrapolated result for the $\theta$-dependence of the free
energy density, both at zero and finite temperature.
Section~\ref{sec:axion} is dedicated to the analysis of the
consequences of our results in the context of axion
cosmology. Finally, in Section~\ref{discussion}, we draw our
conclusions and discuss future perspectives.

\section{Numerical setup}

\label{sec:setup}

\subsection{Discretization adopted}

The action of $N_f=2+1$ QCD is discretized by using the tree level improved
Symanzik gauge action \cite{weisz, curci}  and the stout improved staggered
Dirac operator. Explicitly, the Euclidean partition function is written as
\begin{align}
\mathcal{Z} &= \int \!\mathcal{D}U \,e^{-\mathcal{S}_{Y\!M}} \!\!\!\!\prod_{f=u,\,d,\,s} \!\!\! 
\det{\left({M^{f}_{\textnormal{st}}[U]}\right)^{1/4}}
,  \label{partfunc} \\
\mathcal{S}_{Y\!M}&= - \frac{\beta}{3}\sum_{i, \mu \neq \nu} \left( \frac{5}{6}
W^{1\!\times \! 1}_{i;\,\mu\nu} -
\frac{1}{12} W^{1\!\times \! 2}_{i;\,\mu\nu} \right), \label{tlsyact} \\
(M^f_{\mathrm{st}})_{i,\,j}&=am_f \delta_{i,\,j}+ \nonumber\\ 
&+\sum_{\nu=1}^{4}\frac{\eta_{i;\,\nu}}{2}
\left[U^{(2)}_{i;\,\nu}\delta_{i,j-\hat{\nu}} -U^{(2)\dagger}_{i-\hat\nu;\,\nu}
\delta_{i,j+\hat\nu}  \right] \ . \label{fermmatrix}
\end{align}
The symbol $W^{n\times m}_{i;\ \mu\nu}$ denotes the trace of the
$n\times m$ Wilson loop built using the gauge links departing from the
site in position $i$ along the positive $\mu,\nu$ directions. The
gauge matrices $U^{(2)}_{i,\mu}$, used in the definition of the
staggered Dirac operator, are related to the gauge links $U_{i;\ \mu}$
(used in $\mathcal{S}_{YM}$) by two levels of stout-smearing
\cite{Morningstar:2003gk} with isotropic smearing parameter
$\rho=0.15$.

The bare parameters $\beta$, $m_s$ and $m_l\equiv m_u=m_d$ were chosen
in such a way to have physical pion mass $m_{\pi}\approx
135\,\mathrm{MeV}$ and physical $m_s/m_l$ ratio. The values of the
bare parameters used in this study to move along this line of constant
physics are reported in Tab.~\ref{tab:bareparam}. Most of them were
determined in \cite{physline1, physline2}, the remaining have been
extrapolated by using a cubic spline interpolation. The lattice
spacings reported in Tab.~\ref{tab:bareparam} have a $2-3\%$ of
systematic uncertainty, as discussed in \cite{physline1, physline2}
(see also \cite{physline3}).
 
\begin{table}
\centering
\begin{tabular}{|c|c|c|}
  \hline
  \rule{0mm}{3.2mm} $\beta$ & $a$ [fm] & $a m_s$   \\ \hline
  \rule{0mm}{3.2mm}3.750 & 0.1249 & $5.03\times 10^{-2}$ \\ \hline
  \rule{0mm}{3.2mm}3.850 & 0.0989 & $3.94\times 10^{-2}$ \\ \hline
  \rule{0mm}{3.2mm}3.938 & 0.0824 & $3.30\times 10^{-2}$ \\ \hline
  \rule{0mm}{3.2mm}4.020 & 0.0707 & $2.81\times 10^{-2}$ \\ \hline
  \rule{0mm}{3.2mm}4.140 & 0.0572 & $2.24\times 10^{-2}$ \\ \hline
\end{tabular}
\caption{Bare parameters used in this work, from \cite{physline1,
    physline2} or spline interpolation of data thereof. The
  systematic uncertainty on the lattice spacing determination is
  $2-3\%$ and the light quark mass is fixed by using
  $m_s/m_l=28.15$.} \label{tab:bareparam}
\end{table}

\subsection{Determination of the topological content}

In order to expose the topological content of the gauge
configurations, we adopt a gluonic definition of the topological
charge density, measured after a proper smoothing procedure, which has
been shown to provide results equivalent to definitions based on
fermionic operators~\cite{Neuberger:1997fp,Hasenfratz:1998ri,
  Luscher:1998pqa,Luscher:2004fu,Giusti:2008vb}.  The basic underlying
idea is that, close enough to the continuum limit, the topological
content of gauge configurations becomes effectively stable under a
local minimization of the gauge action, while ultraviolet fluctuations
are smoothed off.

A number of smoothing algorithms has been devised along the time.  A well known 
procedure is cooling~\cite{cooling}: an elementary
cooling step consists in the substitution of each link variable
$U_{i;\ \mu}$ with the $SU(3)$ matrix that minimizes the Wilson action,
in order to reduce the UV noise.  While for the case of the $SU(2)$
group this minimization can be performed analytically, in the $SU(3)$
case the minimization is usually performed \emph{\`a la}
Cabibbo-Marinari, i.e. by iteratively minimizing on the $SU(2)$
subgroups. When this elementary cooling step is subsequently applied
to all the lattice links we obtain a cooling step.

Possible alternatives may consist in choosing a different action to be
minimized during the smoothing procedure, or in performing a continuous
integration of the minimization equations. The latter procedure is known as the
gradient flow~\cite{Luscher:2009eq, Luscher:2010iy} and has been shown to
provide results equivalent to cooling regarding topology~\cite{Bonati:2014tqa,
Cichy:2014qta, Namekawa:2015wua, Alexandrou:2015yba}. Because of its
computational simplicity in this work we will thus use cooling, however it will
be interesting to consider in future studies also the gradient flow, especially
as an independent way to fix the physical scale \cite{Luscher:2010iy,
Sommer:2014mea, Fodor:2014cpa}.

Since the topological charge will be measured on smoothed
configurations, we can use the simplest discretization of the
topological charge density with definite parity
\cite{DiVecchia:1981qi}
\begin{equation}\label{eq:qL}
q_L(x) = -\frac{1}{2^9 \pi^2}\sum_{\mu\nu\rho\sigma = \pm 1}^{\pm 4} 
\tilde{\epsilon}_{\mu\nu\rho\sigma} \tr \left(U_{x;\, \mu\nu} U_{x;\, \rho\sigma} \right) \; ,
\end{equation}
where $U_{x;\, \mu\nu}$ is the plaquette located in $x$ and directed
along the $\mu, \nu$ directions. The tensor $\tilde{\epsilon}$ is the
completely antisymmetric tensor which coincides with the usual
Levi-Civita tensor $\epsilon_{\mu\nu\rho\sigma}$ for positive indices
and is defined by ${\tilde{\epsilon}}_{\mu\nu\rho\sigma} =
-{\tilde{\epsilon}}_{(-\mu)\nu\rho\sigma}$ and antisymmetry for
negative indices.

The lattice topological charge $Q_L=\sum_x q_L(x)$ is not in general
an integer, although its distribution gets more and more peaked around
integer values as the continuum limit is approached.  In order to
assign to a given configuration an integer value of the topological
charge we will follow the prescription introduced in
\cite{DelDebbio:2002xa}: the topological charge is defined as
$Q=\mathrm{round}(\alpha Q_L)$, where `$\mathrm{round}$' denotes the
rounding to the closest integer and $\alpha$ is fixed by the
minimization of
\begin{equation}
\left\langle \left( \alpha\, Q_L - \mathrm{round}\left[\alpha\, Q_L \right]\right)^2\right\rangle\ ,
\end{equation} 
i.e. in such a way that $\alpha Q_L$ is `as integer' as possible.
Actually, one could take the non-rounded definition $Q = Q_L$ as well,
the only difference being a different convergence of results to the
common continuum limit (see Ref.~\cite{Bonati:2015sqt} for a more
detailed discussion on this point).  The topological susceptibility is
then defined by
\begin{equation}\label{eq:chi}
a^4 \chi=\frac{\langle Q^2\rangle}{V_4}\ ,
\end{equation}
where $V_4=N_tN_s^3$ is the four-dimensional volume of the lattice,
and the coefficient $b_2$ has been introduced in Eq.~(\ref{eq:b2}).

We measured the topological charge every $20$ cooling steps up to a
maximum of $120$ steps and verified the stability of the topological
quantities under smoothing.  Results that will be presented in the
following have been obtained by using $100$ cooling steps, a number
large enough to clearly identify the different topological sectors but
for which no significant signals of tunneling to the trivial sector
are distinguishable in mean values.

\begin{figure}
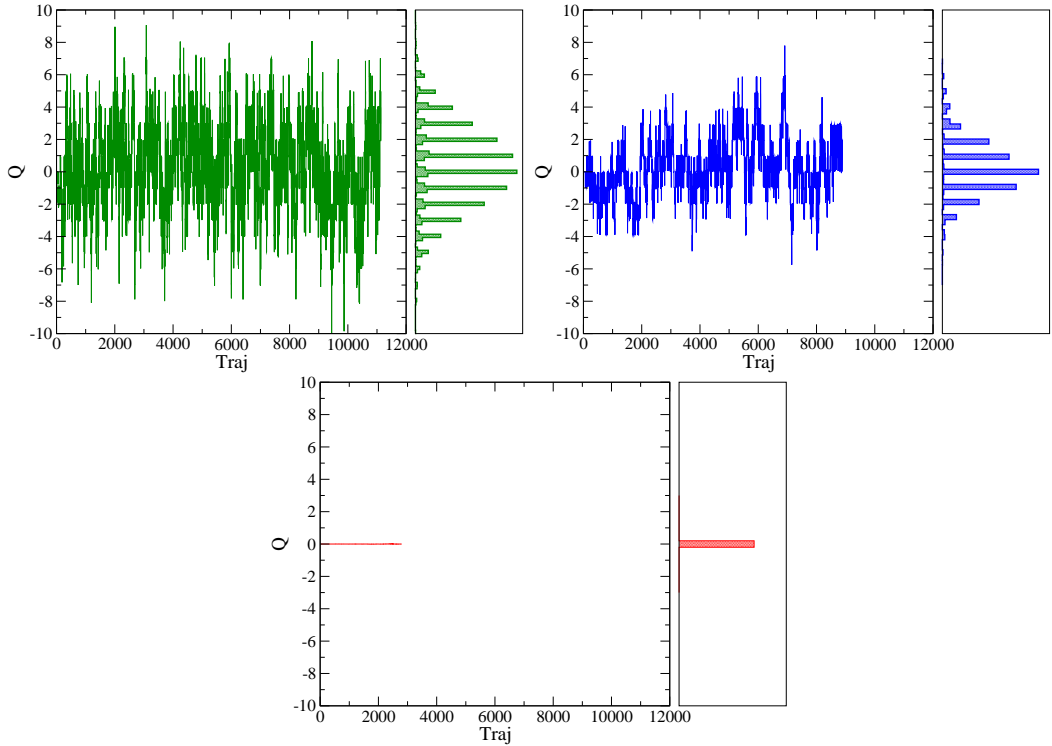

\centering
\includegraphics*[width=0.45\columnwidth]{Q_hist_3.938_L32.eps}
\includegraphics*[width=0.45\columnwidth]{Q_hist_4.140_L48.eps}
\includegraphics*[width=0.45\columnwidth]{Q_hist_4.360_L40.eps}
\caption{Check for the correct sampling of the topological
  charge. (Upper left) Topological charge time history and histogram for
  $a=0.0824\,\mathrm{fm}$ on a $32^4$ lattice.  (Upper right) Topological
  charge time history and histogram for $a=0.0572\,\mathrm{fm}$ on a
  $48^4$ lattice.  (Lower) Topological charge time history for
  $a=0.0397\,\mathrm{fm}$ on a $40^4$ lattice; in this case the run
  was stopped earlier because of the visible freezing of the
  topological charge. Note that the upper histories
  correspond to approximately the same physical volume: hence the
  visible shrink in the topological charge distribution is the actual
  effect of the approach to the continuum limit (see the discussion in
  Sec.~\ref{sec:results}).  }\label{fig:topohistory}
\end{figure}

For all data reported in the following we verified that the
corresponding time histories were long enough to correctly sample the
topological charge. In particular we checked that $\langle Q\rangle$
is compatible with zero and that the topological charge is not frozen.
Indeed it is well known that, while approaching the continuum limit,
the autocorrelation time of the topological charge increases very
steeply until no tunneling events between different sectors happen
anymore~\cite{Alles:1996vn, DelDebbio:2002xa, DelDebbio:2004xh,
Schaefer:2010hu, Kitano:2015fla}.  An example of this behavior can be observed in
Fig.~\ref{fig:topohistory}, where some time-histories for zero
temperature runs are showed, for three different lattice
spacings. While the general features of this phenomenon are common to
all the lattice actions, the critical value of the lattice spacing at
which the charge gets stuck may depend on the specific discretization
adopted. In our case we were able to obtain reasonable sampling for
lattice spacings down to $a=0.0572\,\mathrm{fm}$, while finer lattices
(in particular, with $a \lesssim 0.04$ fm) showed severe freezing over
thousands of trajectories and had to be discarded.

\section{Numerical Results}

\label{sec:results}

The main purpose of our numerical study is to provide results for the
$\theta$-dependence of QCD at finite temperature, in particular above
the pseudo-critical temperature, in order to take them as an input for
axion phenomenology. However, in order to make sure that our lattice
discretization is accurate enough to reproduce the chiral properties
of light fermions, we have also performed numerical simulations at
zero or low temperature, where results can be compared to reliable
analytic predictions.

Indeed, at zero temperature, the full $\theta$ dependence of the QCD
partition function, can be estimated reliably using chiral Lagrangians
\cite{Weinberg:1977ma,DiVecchia:1980ve,Leutwyler:1992yt}.  In
particular, at leading order in the expansion, $\chi$ and $b_2$ can be
written just in terms of $m_\pi$, $f_\pi$ and $z=m_u/m_d$ as
\begin{equation}
\chi^{\rm LO}=\frac{z}{(1+z)^2} m_\pi^2 f_\pi^2\,,
\qquad
b_2^{\rm LO}=-\frac{1}{12}\frac{1+z^3}{(1+z)^3}\,.
\end{equation}
NLO corrections have been computed in \cite{Mao:2009sy, Guo:2015oxa,
diCortona:2015ldu} and are of the order of percent for physical values of the
light quark masses.  Using the estimate in \cite{diCortona:2015ldu} we have
$\chi^{1/4}=75.5(5)$~MeV and $b_2=-0.029(2)$ for $z=0.48(3)$, while
$\chi^{1/4}=77.8(4)$~MeV and $b_2=-0.022(1)$ for $z=1$ used here.  Note that
for pure SU(3) Yang-Mills these numbers become $\chi^{1/4}_{N_f=0}\approx
180$~MeV \cite{Alles:1996nm, DelDebbio:2002xa, DelDebbio:2004ns, Durr:2006ky,
Luscher:2010ik, Cichy:2015jra, Ce:2015qha} and\footnote{It is interesting to
note the striking agreement between the ChPT prediction for $b_2$ in the case
of degenerate light flavors and the numerical results obtained for it in the
quenched theory~\cite{Bonati:2015sqt}.  That seems to suggest a similar form of
$\theta$-dependence in the low temperature phases of the full and of the
quenched theory.  At this stage we have no particular explanation for this
agreement, which might be considered as a coincidence.}
$b_{2,N_f=0}=-0.0216(15)$~\cite{Bonati:2015sqt} (see also
Refs.~\cite{DelDebbio:2002xa,D'Elia:2003gr,Giusti:2007tu,Panagopoulos:2011rb,Ce:2015qha}
for previous determinations in the literature).

ChPT provides a prediction at finite temperature as well. In
particular we have
\begin{equation}\label{tchpt}
\begin{aligned}
\chi_{\rm ChPT}(T) & =\chi(0)\, \left ( 1-\frac32
\frac{T^2}{f_\pi^2}\, J_1 \left [ \frac{T^2}{m_\pi^2} \right ] \right
)\, , \\ b_2^{\rm ChPT}(T) & =b_2(0)\, \left ( 1+ \frac92
\frac{m_\pi^2}{f_\pi^2}\, \frac{z(1+z)}{1+z^3} J_2 \left [
  \frac{T^2}{m_\pi^2} \right ] \right ) \, ,
\end{aligned}
\end{equation}
where the functions $J_n$ are defined in
Ref.~\cite{diCortona:2015ldu}.  However, at temperatures around and
above $T_c$, the chiral condensate drops and chiral Lagrangians break
down, so that the finite $T$ predictions in Eq.~(\ref{tchpt}) are
expected to fail.  In this regime non-perturbative computations based
on first principle QCD are mandatory.

\subsection{Zero Temperature}

\begin{figure}
\centering
\includegraphics[width=0.60\columnwidth, clip]{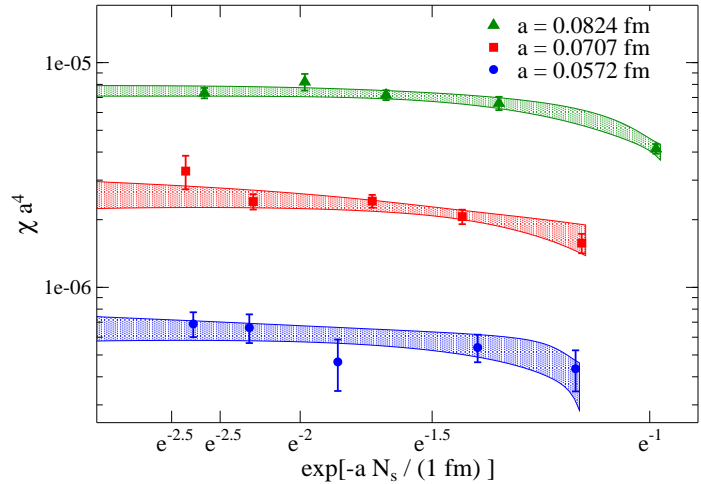}
\caption{Volume dependence of the topological susceptibility for different
lattice spacings. The lattice extent is reported in physical units, while
$\chi$ in lattice spacing units. Bands represent best fits to
Eq.~(\ref{finiteVchi}) for the infinite volume limit
extrapolation.}\label{fig:vol_chi}
\end{figure}

In Tab.~\ref{tab:risT0} we report our determinations of the
topological susceptibility on $N_s^4$ lattices for several values of
$N_s$ and different lattice spacings. The results on the three smallest
lattice spacings are also plotted in Fig.~\ref{fig:vol_chi}.

In order to extract the infinite volume limit of $\chi$ at fixed lattice
spacing, one can either consider only results obtained on the largest available lattices
and fit them to a constant, or try to model the behavior of $\chi$ with the
lattice size using all available data. We have followed both procedures in
order to better estimate possible systematics.

The topological susceptibility can be written as the integral of the
two point function of the topological charge density; since the
$\eta'$ is the lightest intermediate state that significantly
contributes to this two point function (three pions states are OZI
suppressed), one expects the leading asymptotic behavior to be
\begin{equation}
\chi_{N_s}\sim \chi_{\infty}+Ce^{-aN_s m_{\eta'}}\ ,
\label{finiteVchi}
\end{equation}
where $C$ is an unknown constant. This form nicely fits data in the
whole available range of $aN_s$ (see Fig.~\ref{fig:vol_chi}), both
when $m_{\eta'}$ is fixed to its physical value and when it is left as
a fit parameter. The results obtained using this last procedure (which
gives the most conservative estimates) are reported in
Tab.~\ref{tab:risT0} and correspond to the entries denoted by
$N_s=\infty$.  On the other hand, a fit to a constant value works well
when using only data coming from lattices with
$aN_s>1.5\,\mathrm{fm}$, and provides consistent results within
errors.  This analysis makes also us confident that results obtained
on the lattices with $a\simeq 0.1249$~fm and $a \simeq 0.0989$~fm,
where a single spatial volume is available with $a N_s > 3$ fm, are
not affected by significant finite size effects.

\begin{figure}[t!]
\centering
\includegraphics[width=0.60\columnwidth, clip]{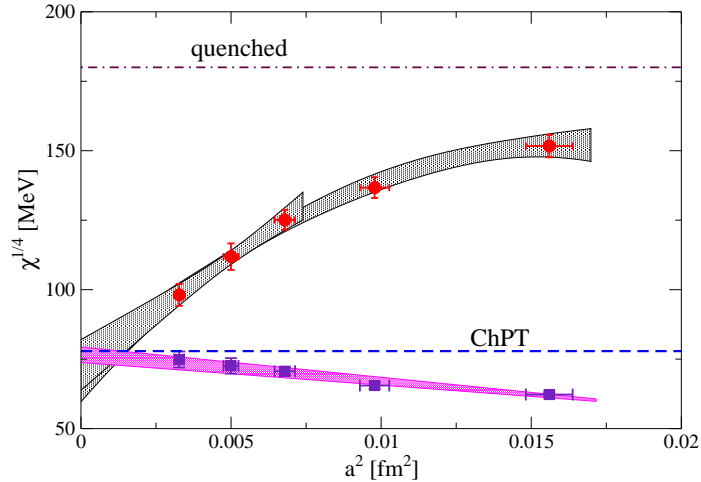}
\caption{Continuum limit of the topological susceptibility: our final result is
$\chi^{1/4}=73(9)\,\mathrm{MeV}$.  Also shown is the value expected from chiral
perturbation theory, denoted by the horizontal dashed line, together with the
quenched value. Square points correspond to the combination 
$\chi^{1/4}_{tc}(a)$ defined in Eq.~\eqref{eq:tastecomb}.}\label{fig:chi_T0}
\end{figure}

In Fig.~\ref{fig:chi_T0} we plot $\chi^{1/4}$, extrapolated to
infinite volume, against the square of the lattice spacing, together
with the ChPT prediction.  Finite cut-off effects are significant,
especially for $a \gtrsim 0.1\,\mathrm{fm}$, meaning that we are not
close enough to the continuum limit to reproduce the correct chiral
properties of light quarks.  In the case of staggered quarks, such
lattice artifacts originate mostly from the fact that the full chiral
symmetry group is reproduced only in the continuum limit, so that the
pion spectrum is composed of a light pseudo-Goldstone boson and other
massive states which become degenerate only as $a \to 0$. The
physical signal vanishes in the chiral limit whereas this is not the
case for its discretization effects. This means that it is necessary
to work on very fine lattice spacings in order to keep these effects
under control, a task which is particularly challenging with present
algorithms, because of the freezing of the topological charge.

Despite the large cut-off effects, we can perform the continuum
extrapolation of our data. If one considers only the three finest
lattice spacings, the finite cut-off effects for this quantity can be
parametrized as simple $O(a^2)$ corrections
\begin{equation}
\chi^{1/4}(a)=\chi^{1/4}+B a^2
\label{chiextrap}
\end{equation}
and a best fit yields $\chi^{1/4} = 73(9)$ MeV, while in order to
describe the whole range of available data one must take into account
$O(a^4)$ corrections as well, i.e.
\begin{equation}
\chi^{1/4}(a)=\chi^{1/4}+B a^2 + C a^4\, ,
\label{chiextrap2}
\end{equation} obtaining 
$\chi^{1/4} = 69(9)$ MeV. Both best fits are reported in
Fig.~\ref{fig:chi_T0}. Therefore we conclude that the continuum
extrapolation is already under control with the available lattice
data, and is in satisfactory agreement with the predictions of chiral
perturbation theory.

In order to further inquire about the reliability of our continuum extrapolation
and the importance of the partial breaking of chiral symmetry in the
staggered discretization, we studied the combination
\begin{equation}\label{eq:tastecomb}
\chi^{1/4}_{tc}(a)=a\chi^{1/4}(a)\frac{m_{\pi}^{\mathrm{phys}}}{a m_{ngb}(a)}\ .
\end{equation}
Here $m_{ngb}(a)$ is the mass of one of the non-Goldstone pions, i.e. of a
state that becomes massless in the chiral limit only if the continuum limit is
performed; since in the continuum $m_{ngb}(a)\to m_{\pi}^{\mathrm{phys}}$, this
ratio converges to $\chi^{1/4}$ as $a\to 0$.  The state with taste structure
$\gamma_i\gamma_{\mu}$ was used, whose mass is close to the root mean square
value of all the taste states (see Ref. \cite{physline2} Fig. 2) and the values
of $\chi^{1/4}_{tc}(a)$ are shown in Fig.~\ref{fig:chi_T0} as square points.
It is clear that the combination $\chi^{1/4}_{tc}(a)$ strongly reduces lattice
artefacts with respect to $\chi^{1/4}(a)$, moreover a linear fit in $a^2$ well
describes data for all available lattice spacings, giving the result
$\chi^{1/4}=77(3)$. Although a complete study of the systematics affecting
$\chi^{1/4}_{tc}(a)$ was not performed (e.g. the dependence of $m_{ngb}(a)$ on
the lattice size was not studied, just the largest size was used), this is a
strong indication that the dominant source of lattice artefacts in
$\chi^{1/4}(a)$ is the chiral symmetry breaking present at finite lattice
spacing in the staggered discretization. 

\begin{table}
\centering
\begin{tabular}{|c|c|c|} 
\hline \rule{0mm}{3.2mm}$N_s$ & $a$ [fm] & $\chi a^4$ \\ \hline
\hline \rule{0mm}{3.2mm}32 & 0.1249 & $8.55(32)\times 10^{-5}$\\ \hline
\hline \rule{0mm}{3.2mm}32 & 0.0989 & $2.22(10)\times 10^{-5}$\\ \hline
\hline \rule{0mm}{3.2mm}12 & 0.0824 & $4.14(20)\times 10^{-6}$ \\
\hline \rule{0mm}{3.2mm}16 & 0.0824 & $6.60(45)\times 10^{-6}$ \\
\hline \rule{0mm}{3.2mm}20 & 0.0824 & $7.17(36)\times 10^{-6}$ \\
\hline \rule{0mm}{3.2mm}24 & 0.0824 & $8.21(70)\times 10^{-6}$ \\
\hline \rule{0mm}{3.2mm}32 & 0.0824 & $7.32(39)\times 10^{-6}$ \\
\hline \rule{0mm}{3.2mm}$\infty$ & 0.0824 & $7.50(40)\times 10^{-6}$ \\ \hline 
\hline \rule{0mm}{3.2mm}16 & 0.0707 & $1.57(16)\times 10^{-6}$ \\
\hline \rule{0mm}{3.2mm}20 & 0.0707 & $2.06(15)\times 10^{-6}$ \\
\hline \rule{0mm}{3.2mm}24 & 0.0707 & $2.42(16)\times 10^{-6}$ \\
\hline \rule{0mm}{3.2mm}32 & 0.0707 & $2.41(19)\times 10^{-6}$ \\ 
\hline \rule{0mm}{3.2mm}40 & 0.0707 & $3.29(56)\times 10^{-6}$ \\
\hline \rule{0mm}{3.2mm}$\infty$ & 0.0707 & $2.60(36)\times 10^{-6}$  \\ \hline 
\hline \rule{0mm}{3.2mm}20 & 0.0572 & $4.34(90)\times 10^{-7}$ \\
\hline \rule{0mm}{3.2mm}24 & 0.0572 & $5.40(76)\times 10^{-7}$ \\
\hline \rule{0mm}{3.2mm}32 & 0.0572 & $4.7(1.2)\times 10^{-7}$ \\
\hline \rule{0mm}{3.2mm}40 & 0.0572 & $6.61(96)\times 10^{-7}$ \\
\hline \rule{0mm}{3.2mm}48 & 0.0572 & $6.88(87)\times 10^{-7}$ \\
\hline \rule{0mm}{3.2mm}$\infty$ & 0.0572 & $6.60(82)\times 10^{-7}$  \\ \hline
\end{tabular}
\caption{Topological susceptibility in lattice units, measured on
  several $N_s^4$ lattices for different values of the lattice
  spacing. Values denoted by $N_s=\infty$ correspond to the
  thermodynamical extrapolation and they are obtained by a fit to
  exponential without fixing $m_{\eta'}$ to its physical value, see
  Eq.~(\ref{finiteVchi}).
}\label{tab:risT0}
\end{table}

\subsection{Finite Temperature}

Finite temperature simulations can in principle be affected by lattice
artifacts comparable to those present at $T = 0$.  For that reason, we
have limited our finite $T$ simulations to the three smallest lattice
spacings explored at $T = 0$, i.e.~those in the scaling window adopted
for the extrapolation to the continuum limit with only $O(a^2)$
corrections. At fixed lattice spacing the temperature has been varied
by changing the temporal extent $N_t$ of the lattice and, in all
cases, we have fixed $N_s=48$. This gives a spatial extent equal or
larger than those explored at $T = 0$ and an aspect ratio $N_s / N_t
\geq 3$ for all explored values of $N_t$.  The absence of significant
finite volume effects has also been verified directly by comparing
results with those obtained on $N_t\times 40^3$ lattices.  In
Tab.~\ref{tab:risT} we report the numerical values obtained for the
topological susceptibility, for the ratio $\chi(T,a)/\chi(T=0,a)$
(where for $\chi(T=0,a)$ the infinite volume extrapolation has been
taken) and for the cumulant ratio $b_2$.

\begin{table}
\centering
\begin{tabular}{|c|c|c|c|c|c|c|}
\hline 
\rule[-2mm]{0mm}{5.5mm} $N_t$ & $a$ [fm] & $T/T_c$  &  $\chi(T,a)a^4$   & $\frac{\chi(T,a)}{\chi(0,a)}$  & $b_2$ \\ 
\hline\hline
\rule{0mm}{3.2mm}16  & 0.0824 &  0.964 & $6.12(48)\cdot 10^{-6}$ &  $0.817(75)$  & -0.23(33)  \\ \hline
\rule{0mm}{3.2mm}14  & 0.0824 &  1.102 & $5.25(47)\cdot 10^{-6}$ &  $0.701(70)$  & -0.191(95) \\ \hline
\rule{0mm}{3.2mm}12  & 0.0824 &  1.285 & $4.04(11)\cdot 10^{-6}$ &  $0.539(29)$  & -0.003(25)  \\ \hline
\rule{0mm}{3.2mm}10  & 0.0824 &  1.542 & $2.87(12)\cdot 10^{-6}$ &  $0.383(23)$  & -0.083(29) \\ \hline
\rule{0mm}{3.2mm}8   & 0.0824 &  1.928 & $1.296(97)\cdot 10^{-6}$&  $0.173(15)$  & -0.065(25) \\ \hline \hline
\rule{0mm}{3.2mm}24  & 0.0707 &  0.749 & $2.82(36)\cdot 10^{-6}$ &  $1.08(18)$   & -0.11(12)  \\ \hline
\rule{0mm}{3.2mm}16  & 0.0707 &  1.124 & $1.53(15)\cdot 10^{-6}$ &  $0.589(84)$  & -0.057(43) \\ \hline
\rule{0mm}{3.2mm}14  & 0.0707 &  1.284 & $1.385(79)\cdot 10^{-6}$ & $0.532(62)$  & -0.019(30) \\ \hline
\rule{0mm}{3.2mm}12  & 0.0707 &  1.498 & $1.05(11)\cdot 10^{-6}$ &  $0.403(60)$  & -0.101(21) \\ \hline
\rule{0mm}{3.2mm}10  & 0.0707 &  1.798 & $7.87(87)\cdot 10^{-7}$ &  $0.302(46)$  & -0.096(26) \\ \hline
\rule{0mm}{3.2mm}8   & 0.0707 &  2.247 & $3.41(31)\cdot 10^{-7}$ &  $0.131(18)$  & -0.077(18) \\ \hline
\rule{0mm}{3.2mm}6   & 0.0707 &  2.996 & $1.22(21)\cdot 10^{-7}$ &  $0.0469(94)$ & -0.076(10) \\ \hline \hline
\rule{0mm}{3.2mm}16  & 0.0572 &  1.389 & $2.92(41)\cdot 10^{-7}$ &  $0.41(16)$  & -0.049(17) \\ \hline 
\rule{0mm}{3.2mm}14  & 0.0572 &  1.587 & $2.04(32)\cdot 10^{-7}$ &  $0.29(11)$  & -0.058(16) \\ \hline
\rule{0mm}{3.2mm}12  & 0.0572 &  1.852 & $1.43(19)\cdot 10^{-7}$ &  $0.202(79)$  & -0.0626(90) \\ \hline
\rule{0mm}{3.2mm}10  & 0.0572 &  2.222 & $8.3(1.4)\cdot 10^{-8}$ &  $0.118(48)$  & -0.0705(74) \\ \hline
\rule{0mm}{3.2mm}8   & 0.0572 &  2.777 & $4.85(86)\cdot 10^{-8}$ &  $0.068(27)$  & -0.091(11) \\ \hline
\rule{0mm}{3.2mm}6   & 0.0572 &  3.703 & $2.51(72)\cdot 10^{-8}$ &  $0.035(16)$  & -0.0792(12) \\ \hline
\end{tabular}
\caption{Finite temperature results; in all cases $N_s=48$ was used.}\label{tab:risT}
\end{table}

\begin{table}
\centering
\begin{tabular}{|c|c|c|c|}
\hline\rule{0mm}{3.2mm}$T_{cut}/T_c$  & $A_0$ (MeV)    & $A_1$(fm$^{-2}$)   & $A_2$ \\ \hline
\rule{0mm}{3.2mm}1.2  &  76.0(5.1)  & 103(17)  & -0.671(40) \\ \hline
\rule{0mm}{3.2mm}1.4  &  80.2(6.9)  & 100(20)  & -0.728(57) \\ \hline
\rule{0mm}{3.2mm}1.6  &  86(10)     & 85(24)   & -0.750(80) \\ \hline
\rule{0mm}{3.2mm}2.0  &  83(18)     & 97(53)   & -0.718(14) \\ \hline
\end{tabular}
\caption{Best fit values for the coefficients in 
\eqref{eq:fit_susc} obtained by fitting only data corresponding to temperatures
$T>T_{cut}$.}\label{tab:fit_susc}
\end{table}

\begin{figure}[tbh]
\centering
\includegraphics[width=0.60\columnwidth, clip]{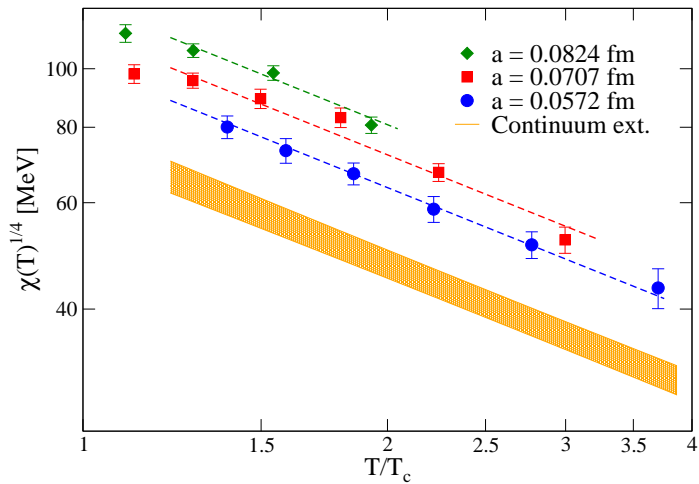}
\caption{Continuum extrapolation of the fourth root topological susceptibility,
using the function $\chi^{1/4}(a,T)=A_0(1+A_1 a^2)(T/T_c)^{A_2}$. Only data
corresponding to $T\gtrsim 1.2T_c$ have been used in the fit.} \label{fig:chi_T}
\end{figure}

\begin{table}
\centering
\begin{tabular}{|c|c|c|c|}
\hline\rule{0mm}{3.2mm}$T_{cut}/T_c$  & $D_0$      & $D_1$(fm$^{-2}$)    & $D_2$ \\ \hline
\rule{0mm}{3.2mm}1.2            & 1.17(24)   & -2(28)  & -2.71(15) \\ \hline
\rule{0mm}{3.2mm}1.4            & 1.56(39)   & -12(30)  & -3.02(23) \\ \hline
\rule{0mm}{3.2mm}1.6            & 1.81(76)   & -33(32)  & -2.99(39) \\ \hline
\rule{0mm}{3.2mm}2.0            & 1.8(1.5)   & 2(22)$\times 10^3$  & -2.90(65) \\ \hline
\end{tabular}
\caption{Best fit values for the coefficients in
\eqref{eq:fit_suscratio} obtained by fitting only data corresponding to
temperatures $T>T_{cut}$.}\label{tab:fit_suscratio}
\end{table}
Results for $\chi^{1/4}$ as a function of $T/T_c$, where $T_c = 155$ MeV, are
reported in Fig.~\ref{fig:chi_T} for the three different lattice spacings.  The
dependence on $a$ is quite strong, as expected from the $T = 0$ case.  Inspired
by the instanton gas prediction, Eq.~(\ref{eq:chi_inst_pert}), we have
performed a fit with the following ansatz
\begin{equation}\label{eq:fit_susc}
\chi^{1/4}(a,T)=A_0(1+A_1 a^2)\left(\frac{T}{T_c}\right)^{A_2}\ ,
\end{equation}
which also takes into account the dependence on $a$ and nicely describes all
data in the range $T > 1.2\, T_c$ with $\chi^2/\mathrm{dof} \simeq 0.7$.  In
Tab.~\ref{tab:fit_susc} we report the best fit values obtained performing the
fit in the region $T>T_{cut}$ for some $T_{cut}$ values; best fit curves are
reported in Fig.~\ref{fig:chi_T} as well, together with a band corresponding to
the continuum extrapolation.

\begin{figure}[tbh]
\centering
\includegraphics[width=0.60\columnwidth, clip]{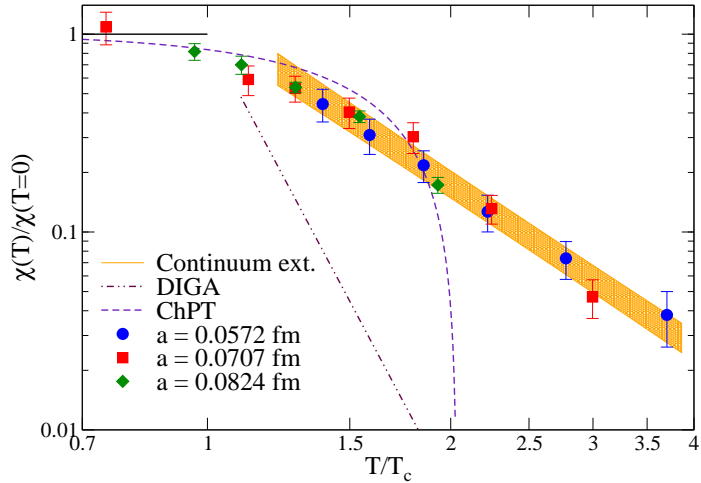}
\caption{Ratio of the topological susceptibilities $\chi(T)/\chi(T=0)$,
evaluated at fixed lattice spacing. The horizontal solid line describes trivial
scaling $\chi(T)=\chi(T=0)$, while the dashed line is the prediction from
finite temperature ChPT and the dashed-dotted line shows the slope predicted by
DIGA computation. The band corresponds to the continuum extrapolation using the
function $\chi(a,T)/\chi(a,T=0)=D_0(1+D_1 a^2)(T/T_c)^{D_2}$, only data
corresponding to $T\gtrsim 1.2T_c$ have been used in the fit.}\label{fig:chi_ratio_T}
\end{figure}

\begin{figure}[tbh]
\centering
\includegraphics[width=0.60\columnwidth, clip]{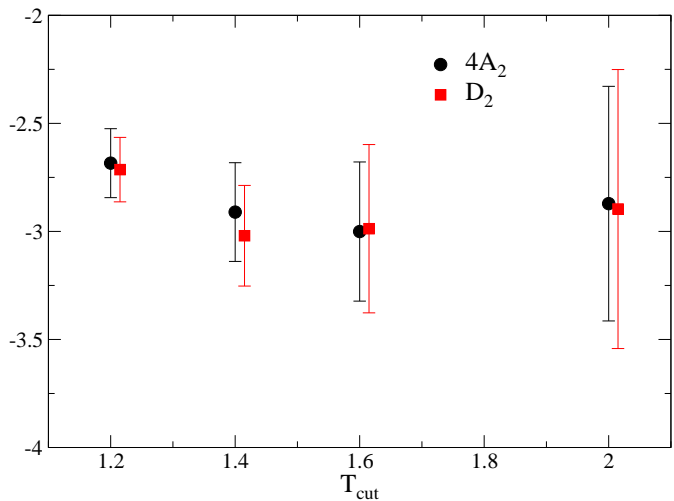}
\caption{Plot of the exponent of the power-law behavior of $\chi(T)$,
extracted from fit to Eqs.~\eqref{eq:fit_susc} and \eqref{eq:fit_suscratio} for
$T>T_{cut}$ (see Tabs.~\ref{tab:fit_susc}-\ref{tab:fit_suscratio}). The value
predicted by the dilute instanton gas approximation and perturbation theory is
$4A_2,D_2 \sim -8$.}\label{fig:power}
\end{figure}

\begin{figure}[tbh]
\centering
\includegraphics[width=0.60\columnwidth, clip]{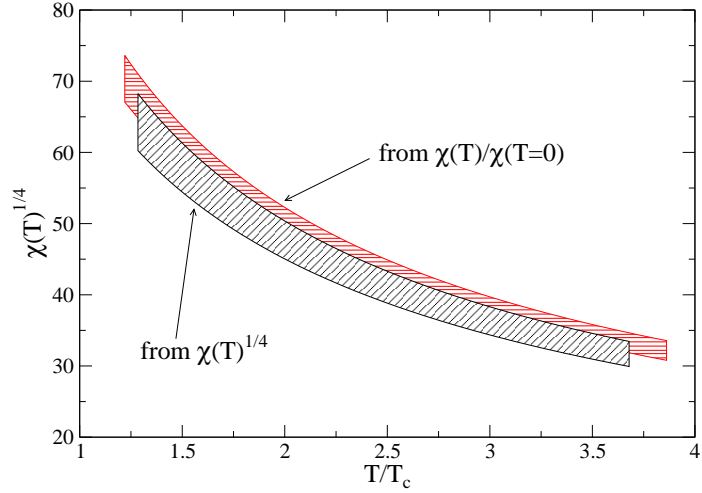}
\caption{Comparison of the two continuum extrapolations for $\chi(T)^{1/4}$.
Continuum extrapolated results for $\chi(T)/\chi(0)$ have been multiplied by
the ChPT prediction for $\chi(0)$.  } \label{confronto}
\end{figure}

\begin{figure}[tbh]
\centering
\includegraphics[width=0.60\columnwidth, clip]{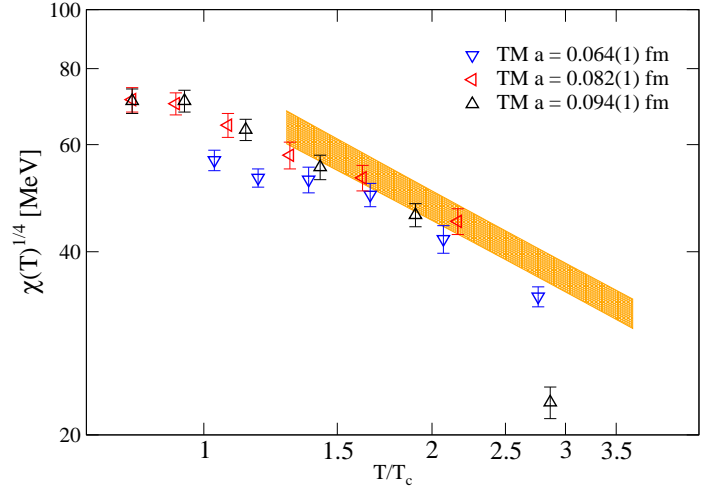}
\caption{Comparison of our continuum extrapolated $\chi^{1/4}(T)$ (band) with
results from Ref.~\cite{Trunin:2015yda}.  Data for $\chi(T)$ obtained at
unphysical quark masses have been rescaled according to the DIGA relation
$\chi(T)\sim m_q^2\sim m_{\pi}^4$.} \label{confronto_bis}
\end{figure}

\begin{figure}[tbh]
\centering
\includegraphics[width=0.60\columnwidth, clip]{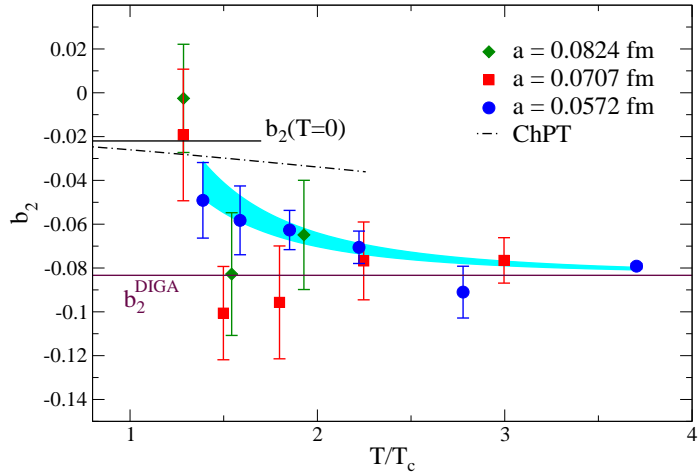}
\caption{$b_2$ evaluated at three lattice spacing. The horizontal solid lines
correspond to the value of $b_2$ at $T=0$ predicted by ChPT (which is
about $-0.022$) and to the instanton-gas expected value $b_2^{DIGA}=-1/12$.
The dotted-dashed line is the prediction of finite temperature ChPT, while
the light blue band is the result of a fit to the smallest lattice spacing data
using Eq.~\eqref{eq:b2_virial} with $d_2$ independent of $T$.}\label{fig:b2} 
\end{figure}

It is remarkable that most of the lattice artifacts disappear when one
considers, in place of $\chi$ itself, the ratio
$\chi(T,a)/\chi(T=0,a)$, whose dependence on the lattice spacing is
indeed quite mild. That is clearly visible in
Fig.~\ref{fig:chi_ratio_T}. Also in this case we have adopted a fit
ansatz similar to Eq.~(\ref{eq:fit_susc})
\begin{equation}\label{eq:fit_suscratio}
\frac{\chi(a,T)}{\chi(a,T=0)}=D_0(1+D_1 a^2) \left(\frac{T}{T_c}\right)^{D_2}\ , 
\end{equation}
which well fits all data in the range $T > 1.2\, T_c$ with a
$\chi^2/\mathrm{dof}\simeq 1$.  The best fit curves and the continuum
extrapolation are shown in Fig.~\ref{fig:chi_ratio_T}.  The best fit
parameters, again for different $T_{cut}$ values, are reported in
Tab.~\ref{tab:fit_suscratio}.  It is important to note, in order to assess the
reliability of our continuum limit, that the two different extrapolations,
Eqs.~(\ref{eq:fit_susc}) and (\ref{eq:fit_suscratio}), provide perfectly
consistent results: that can be appreciated both from Fig.~\ref{fig:power},
where we compare the two coefficients $D_2$ and $4 A_2$, and from
Fig.~\ref{confronto}, where we directly compare the two continuum
extrapolations. As a further check, we have also verified that by
fitting $[\chi(a,T)/\chi(a,T=0)]^{1/4}$ we obtain results in perfect agreement
with the ones obtained by fitting $\chi(a,T)/\chi(a,T=0)$. As an example, using
in the fit the values corresponding to $T>1.2T_c$ one obtains for the exponent
the value $-0.674(38)$, to be compared with $D_2/4$ from the first line of
Tab.~\ref{tab:fit_suscratio} and $A_2$ from the first line of
Tab.~\ref{tab:fit_susc}.  However in the following analysis we will refer to
results obtained through the ratio $\chi(T)/\chi(0)$, which is the quantity
showing smaller finite cut-off corrections.

Let us now comment on our results for the topological susceptibility.
For temperatures below or around $T_c$, the temperature dependence is
quite mild and, for temperatures up to $T \simeq 1.2\, T_c$, even
compatible with the prediction from ChPT, which is reported in
Fig.~\ref{fig:chi_ratio_T} for comparison. Then, for higher values of
$T$, a sharp power law drop starts.  However, it is remarkable that
the power law exponent is smaller, by more than a factor two, with
respect to the instanton gas computation,
Eq.~\eqref{eq:chi_inst_pert}, which predicts $D_2 \simeq -8$ in the
case of three light flavors (the dependence on the number of flavor
is however quite mild).  DIGA is expected to provide the correct
result in a region of asymptotically large temperatures, which however
seems to be quite far from the range explored in the present study,
which goes up to $T \simeq 4\, T_c$.

This is in sharp contrast with the quenched theory, where the DIGA
power law behavior sets in at lower
temperatures~\cite{Borsanyi:2015cka,Berkowitz:2015aua}. In order to
allow for a direct comparison, we have reported the DIGA prediction in
Fig.~\ref{fig:chi_ratio_T}, after fixing it by imposing
$\chi(T_c)=\chi(0)$ as an overall normalization.  As we discuss in the
following section, the much milder drop of $\chi$ as a function of $T$
has important consequences for axion cosmology.

A determination of the topological susceptibility has been presented
in Ref.~\cite{Buchoff:2013nra}, based on the Domain Wall
discretization, in the range $T/T_c \sim $ 0.9 - 1.2. Since the data
have been produced with different lattice spacings at different
temperatures, it is however difficult to compare their results with
ours. An extended range of $T$ has been explored in
Ref.~\cite{Trunin:2015yda} in the presence of twisted mass Wilson
fermions, reporting a behavior similar to that observed in this study,
although with larger values of the quark masses (corresponding to a
pion mass $m_\pi \sim 370$~MeV). The comparison is performed in
Fig.~\ref{confronto_bis}, and shows a reasonable agreement if results
from Ref.~\cite{Trunin:2015yda} are rescaled according to the mass
dependence expected from DIGA\footnote{One might wonder why DIGA
  should work for the mass dependence of $\chi$ and not for its
  dependence on $T$. A possible explanation is that, while the
  temperature dependence stems from a perturbative computation, the
  mass dependence comes from the existence of isolated zero modes in
  the Dirac operator, i.e. from the very hypothesis of the existence
  of a dilute gas of instantons, which seems to be verified already at
  moderately low values of $T$, see the following discussion on $b_2$,
  but not at $T\sim T_c$.}, i.e. $\chi(T)\sim m_q^2\sim m_{\pi}^4$.
\\

Let us now turn to a discussion of our results for the coefficient
$b_2$ (defined in Eq.~\eqref{stheta}), which is related to the
non-gaussianity of the topological charge distribution and gives
information about the shape of the $\theta$-dependent part of the free
energy density. Data for $b_2$ are reported in Fig.~\ref{fig:b2}. For
$T<T_c$ this observable is too noisy to give sensible results but a
reasonable guess, motivated by what happens in the quenched case and
by the ChPT prediction, is that $b_2$ is almost temperature
independent for $T\lesssim T_c$, like the topological susceptibility.
In the high temperature region $b_2$ can instead be measured with
reasonable accuracy and, due to the peculiar dependence of the noise
on this observable on $\chi {\cal V}$, data on the smaller lattice
spacing turned out to be significantly more precise than the others
(see the discussion in Ref.~\cite{Bonati:2015sqt}). While data clearly
approach the dilute instanton gas prediction $b_2^{DIGA}=-1/12$ at
high temperature, deviations from this value are clearly visible on
all the lattice spacings for $T\simeq 1.3~T_c$ and, for the smallest
lattice spacing, also up to $T\sim 2.5~T_c$.

This is in striking contrast to what is observed in pure gauge theory,
where deviations from $-1/12$ are practically absent for $T\gtrsim
1.15~T_c$, with a precision higher than $10\%$ \cite{Bonati:2013tt,
  Borsanyi:2015cka, Bonati:2015uga, Xiong:2015dya}. As discussed in
the introduction deviations from $b_2^{DIGA}$ cannot be simply
ascribed to a failure of perturbation theory (like e.g. for the
behavior of $\chi(T)$) but are instead unambiguous indications of
interaction between instantons. Another difference with respect to the
quenched case is that in the pure gauge theory (with both gauge groups
$SU(N)$ and $G_2$, see \cite{Bonati:2013tt, Bonati:2015uga}) the
asymptotic value is approached from below, while in the present case
it is approached from above. These peculiar features can be related
to a different interaction between instantons mediated by light
quarks, as it is clear from the following discussion.

To describe deviations from the instanton gas behavior it is convenient to use
the parametrization of $\mathcal{F}(\theta,T)$ introduced in
Ref.~\cite{Bonati:2013tt}:
\begin{equation}\label{eq:virial}
\mathcal{F}(\theta, T)=\sum_{n>0}c_{2(n-1)}(T)\sin^{2n}(\theta/2)\ .
\end{equation}
Indeed it is not difficult to show that every even function of $\theta$ of
period $2\pi$ can be written in this form, and the main advantage of this form
is that the value of coefficient $c_{2n}$ influences only $b_{2j}$ with $j\ge
n$: in particular
\begin{equation}
\begin{aligned}
&\chi(T)=c_0(T)/2;\quad b_2(T)=-\frac{1}{12}+\frac{c_2(T)}{8\chi(T)}; \\
&b_4(T)=\frac{1}{360}-\frac{c_2(T)}{48\chi(T)}+\frac{c_4(T)}{32\chi(T)}\ .
\end{aligned}
\end{equation}
Since the coefficients $c_{2n}$ parametrize deviations from the instanton gas
that manifest themselves only in higher-cumulants of the topological charge, it
is natural to interpret Eq.~\eqref{eq:virial} as a virial expansion, in which
the role of the ``density'' is played by the first coefficient (i.e.  the
topological susceptibility $\chi$), and to introduce the dimensionless
coefficients $d_{2n}$ by
\begin{equation}
c_{2(n-1)}(T)=d_{2(n-1)}(T)\frac{\chi(T)^{n}}{\chi(T=0)^{n-1}} \ ,
\end{equation}	
where $\chi(T=0)$ was used just as a dimensional normalization and 
one expects a mild dependence of $d_{2(n-1)}(T)$ on the temperature, 
since the strongly dependent component $\chi(T)^n$ have been factorized.
To lowest order we thus have
\begin{equation}\label{eq:b2_virial}
b_2(T)=-\frac{1}{12}+\frac{d_2}{8}\frac{\chi(T)}{\chi(T=0)}\ ,
\end{equation}
which nicely describes the $b_2$ data for the smallest lattice 
spacing using $d_2=0.80(16)$, see Fig.~\ref{fig:b2}, where 
the expression $\chi(T)/\chi(T=0)\simeq (T_c/T)^{2.7}$ was used.

In the spirit of a virial expansion interpretation of Eq.~(\ref{eq:virial}),
the coefficient $d_2$ can be considered as the lowest order interaction term
between instantons. In particular, a positive value of $d_2$ corresponds to an
attractive potential, which is in sharp contrast with the pure gauge case,
where a repulsive, negative value of $d_2$ is observed. This peculiar
difference can be surely interpreted in terms of effective instanton
interactions mediated by light quarks, which are likely also responsible for
the much slower convergence, with respect to the quenched case, to the DIGA
prediction.

\section{Implications for Axion Phenomenology} \label{sec:axion}

The big departure of the results for the topological susceptibility at finite
temperature from the DIGA prediction has a strong impact on the computation of
the axion relic abundance. In particular the model independent contribution
from the misalignment mechanism is determined to be \cite{diCortona:2015ldu}
\begin{equation}
\Omega_a^{mis}=\frac{86}{33} \frac{\Omega_\gamma}{T_\gamma }  \frac{n_a^\star}{s^\star}m_a\,,
\end{equation}
where $\Omega_\gamma$ and $T_\gamma$ are the present abundance and temperature
of photons while $n_a^\star/s^\star$ is the ratio between the comoving number
density $n_a=\langle m_a a^2\rangle$ and the entropy density $s$ computed at a late time
$t_\star$ such that the ratio $n_a/s$ became constant. The number
density $n_a$ can be extracted by solving the axion equation of motion 
\begin{equation} \label{eq:evo}
\ddot{a} + 3 H \dot{a} + V'(a)=0\, ,
\end{equation}
The temperature (and time) dependence of the Hubble parameter $H$ is determined
by the Friedmann equations and the QCD equation of state.  The biggest
uncertainties come therefore from the temperature dependence of the axion
potential $V(a)$.  At high temperatures the Hubble friction dominates over the
vanishing potential and the field is frozen to its initial value $a_0$.  As the
Universe cools the pull from the potential starts winning over the friction
(this happens when $T\approx T_{osc}$, defined as $m_a(T_{osc})\approx 3
H(T_{osc})$) and the axion starts oscillating around the minimum. Shortly after
$H$ becomes negligible and the mass term is the leading  scale in
Eq.~\eqref{eq:evo}.  In this regime the approximate  WKB solution has the
form 
\begin{equation}\label{eq:wkb} 
\begin{aligned}
a(t) &\sim A(t) \cos\left( \int_{t_0}^t dt^\prime m_a (t^\prime) \right)=  \\  
&= a(t_0) \, \left(\frac{m_a(t_0) R^3(t_0)}{m_a(t) R^3(t)}\right)^{1/2} \, 
\cos\left( \int_{t_0}^t dt^\prime m_a (t^\prime) \right) \, ,  
\end{aligned}
\end{equation}
where $R(t)$ is the cosmic scale factor.  Since the energy density is given by
$\rho_a \sim m_a^2  A^2/2$,  the solution (\ref{eq:wkb}) implies that  what is
conserved in the comoving volume is not the energy density but $N_a  = \rho_a
R^3 /m_a$, which can be interpreted as the number of  axions
\cite{Preskill:1982cy, Abbott:1982af, Dine:1982ah}.  Through the conservation
of the comoving entropy $S$, it follows that  $n_a^\star/s^\star$  becomes an
adiabatic invariant. Hence, it is enough to integrate the equation of motion
\eqref{eq:evo} in the small window around the time when $T\approx   T_{osc}$.
We integrated numerically Eq.~\eqref{eq:evo} in the interval between the time
when $m_a=H/10$ to  that corresponding to $m_a=2400H$ and extract the ratio
$n_a^\star/s^\star$  when $m_a \sim 300 H$, namely a factor a hundred since the
oscillation regime begins.   The value for $T_{osc}$ varies from $T_c$ to
several GeV depending on the axion decay parameter $f_a$ and the temperature
dependence of the axion potential. More details about this standard computation
can be found for example in \cite{Wantz:2009it,diCortona:2015ldu}. In order to
estimate the uncertainty  in the results given below we  varied the  fitting
parameters of the topological susceptibility $D_2$, $D_0$ and  those relative
to the QCD equation of state~\cite{physline2} within the quoted  statistical
and systematic errors. 

\begin{figure}[tbh]
\centering
\includegraphics[width=0.60\columnwidth, clip]{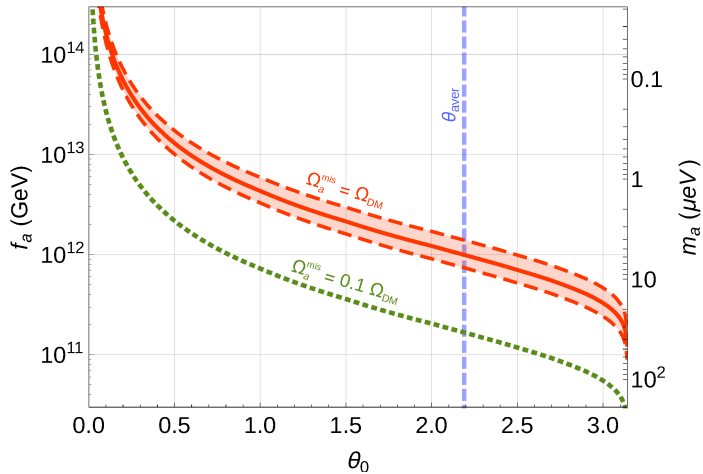}
\caption{\label{fig:favstheta} Values of the axion decay constant $f_a$ as a
function of the initial field value $\theta_0=a_0/f_a$ such that the axion
misalignment contribution matches the full or a tenth of the observed dark
matter abundance (red band or dotted green line respectively). When the PQ
symmetry is broken only after inflation the axion abundance is reproduced by
choosing $\theta_0\approx 2.2$, i.e. the vertical blue dashed line.}
\end{figure}

Given that $b_2(T)$ converges relatively fast to the value predicted
by a single cosine potential, we can assume $V(a)=-\chi(T)\cos(a/f_a)$
for $T\gtrsim T_c$.  
Using the most conservative results for the
fit of $\chi(T)$, i.e.  $\chi(T)/\chi(0)=(1.8\pm 1.5)(T_c/T)^{2.90\pm
  0.65}$, in Fig.~\ref{fig:favstheta} we plot the prediction for the
parameter $f_a$ as a function of the initial value of the axion field
$\theta_0=a_0/f_a$ assuming that the misalignment axion contribution
make up for the whole observed dark matter abundance, $\Omega_{\rm
  DM}=0.259(4)$ \cite{Ade:2015xua}.  We also plot the case where the
axion misalignment contribution accounts only for part (10\% for
definiteness) of the dark matter abundance.

In some cases the axion field acquires all possible values within the visible
horizon, therefore the initial condition to the Eq.~\eqref{eq:evo} needs to be
integrated over. This happens if the PQ symmetry is broken only after inflation
or if the PQ symmetric phase is temporarily restored after inflation (e.g. if
the Hubble scale during inflation or the maximum temperature at reheating are
at or above the PQ scale $f_a$).  Numerically it corresponds to choosing
$\theta_0\approx 2.2$ and the value of the PQ scale in this case is fixed to:
\begin{equation} \label{eq:fasol}
f_a(\theta_0)=(1.00^{+0.40}_{-0.26}{}^{+0.07}_{-0.18}\pm 0.06) \cdot 10^{12}~{\rm GeV}\,,
\end{equation}
where the errors correspond respectively to the uncertainties on the fit
parameters $D_2$, $D_0$, and to the error in the QCD equation of
state.  Note that, in this case, also other contributions from
topological defects are expected to contribute.  In case their effects
are not negligible, or if axions are not responsible for the whole
dark matter abundance, the value above should just be read as an upper
bound to the PQ scale.

The value in Eq.~(\ref{eq:fasol}) is almost one order of magnitude
bigger than the one computed with instantons techniques ($f_a\lesssim
0.2\cdot 10^{12}$~GeV), which in fact corresponds to the value we
instead get for one tenth of the dark matter abundance.

Some few remarks are in order. While the uncertainties on the axion
mass fit we used are not small, the final axion abundance is rather
insensitive to them and the prediction for $f_a$ has therefore a good
precision. The results above, however, rely on the extrapolation of
the axion mass fit formula up to few GeV, where no lattice data is
available.  In particular for the value of $f_a$ in
Eq.~(\ref{eq:fasol}) the axion field starts oscillating around
$T_{osc}=4.3$~GeV. An even longer extrapolation is required for
$f_a=1.67\cdot 10^{11}$~GeV, corresponding to $\Omega_a^{\rm
  mis}=0.1\,\Omega_{\rm DM}$, where the axion starts oscillating
around $T_{osc}=7.2$~GeV.

Given the stability of the fit in the accessible window of
temperatures (see Fig.~\ref{fig:power}) big changes in the axion mass
behavior are not expected.  Still, extending the analysis to even
higher temperatures would be extremely useful to control better the
extrapolation systematics.

Larger values of $f_a$ corresponds instead to smaller $T_{osc}$, for
example for $f_a=10^{13}$~GeV, $T_{osc}=2.2$~GeV while for
$f_a=10^{14}$~GeV, $T_{osc}=1.1$~GeV.  Above these values no
extrapolation is required and the corresponding results are free from
the extrapolation uncertainties.

\section{Conclusions}
\label{discussion}

We studied the topological properties and the $\theta$-dependence
for $N_f = 2+1$ QCD along a line of constant physics, corresponding to
physical quarks masses, and for temperatures up to $4\, T_c$, where $T_c
= 155$ MeV is the pseudo-critical temperature at which chiral symmetry
restoration takes place.  We explored several lattice spacings,
in a range $0.05 - 0.12$ fm, in order to perform a continuum
extrapolation of our results. Our investigation at even smaller
lattice spacings has been hindered by a severe slowing down in the
decorrelation of the topological charge.

At zero temperature we observe large cut-off effects for the
topological susceptibility. Nevertheless we are able to perform a
continuum extrapolation, obtaining from the three finest lattice
spacings $\chi^{1/4} = 73(9)$~MeV, in reasonable agreement with the
ChPT prediction in the case of degenerate light flavors,
$\chi^{1/4}_{\rm ChPT}=77.8(4)$~MeV.

At finite temperature we observe that cut-off effects are drastically
reduced when one considers the ratio $\chi(T) / \chi(T = 0)$, which
turns out to be the most convenient quantity to perform a continuum
extrapolation. The agreement with ChPT persists up to around $T_c$. At
higher temperatures the topological susceptibility presents the
power-law dependence $\chi(T)/\chi(0) = D_0 (T/T_c)^{D_2}$, with $D_2
\sim -3$. The exponent of the power-law behavior is however
substantially smaller than the DIGA prediction, $D_2 \sim - 8$.

Regarding the shape of the $\theta$-dependent part of the free energy
density, and in particular the lowest order coefficient $b_2$, a
visible convergence to the instanton gas prediction is observed in the
explored range. With respect to the pure gauge case, the convergence
of $b_2$ to DIGA is slower and the deviation is opposite in sign
~\cite{Bonati:2013tt, Bonati:2015uga}. This suggests a different
interaction between instantons right above $T_c$, namely repulsive in
the quenched case and attractive in full QCD~\cite{Callan:1977gz}.

The deviations from the dilute instanton gas predictions that we found
in the present study have a significant impact on axions, resulting in
particular in a shift of the axion dark matter window by almost one
order of magnitude with respect to estimates based on DIGA. The softer
temperature dependence of the topological susceptibility also changes
the onset of the axion oscillations, which would now start at a higher
temperatures ($T\sim 4$~GeV).

An important point is that this seems an effect directly related to
the presence of light fermionic degrees of freedom: indeed, pure gauge
computations~\cite{Borsanyi:2015cka,Berkowitz:2015aua} observe a
power-law behavior in good agreement with DIGA in a range of
temperatures comparable to those explored in the present study.

One might wonder whether a different power law behavior might set in
at temperatures higher than 1 GeV. That claims for future studies
extending the range explored by us. The main obstruction to this
extension is represented by the freezing of the topological modes at
smaller lattice spacings which would be necessary to investigate such
temperatures ($a < 0.05$ fm). Such an obstruction could be overcome by
the development of new Monte-Carlo algorithms. Proposals in this
respect have been made in the past~\cite{deForcrand:1997fm,openbc}
and some new strategies have been put forward quite
recently~\cite{surfing, endres}. In view of the exploration of higher
temperatures, one should also consider the inclusion of dynamical
charm quarks.

\acknowledgments
Numerical simulations have been performed on the Blue Gene/Q Fermi
machine at CINECA, based on the agreement between INFN and CINECA
(under INFN project NPQCD). Work partially supported by the INFN SUMA
Project, by the ERC-2010 DaMESyFla Grant Agreement Number: 267985, by
the ERC-2011 NEWPHYSICSHPC Grant Agreement Number: 27975 and by the
MIUR (Italy) under a contract PRIN10-11 protocollo 2010YJ2NYW. FN
acknowledges financial support from the INFN SUMA project.


\begin{thebibliography}{99}

\bibitem{Peccei:1977hh} 
  R.~D.~Peccei and H.~R.~Quinn,
  Phys.\ Rev.\ Lett.\  {\bf 38}, 1440 (1977).
\bibitem{Peccei:1977ur} 
  R.~D.~Peccei and H.~R.~Quinn,
  Phys.\ Rev.\ D {\bf 16}, 1791 (1977).
\bibitem{Wilczek:1977pj}
  F.~Wilczek,
  Phys.\ Rev.\ Lett.\  {\bf 40} (1978) 279.
\bibitem{Weinberg:1977ma}
  S.~Weinberg,
  Phys.\ Rev.\ Lett.\  {\bf 40} (1978) 223.
\bibitem{Preskill:1982cy}
  J.~Preskill, M.~B.~Wise and F.~Wilczek,
  Phys.\ Lett.\ B {\bf 120} (1983) 127.
\bibitem{Abbott:1982af}
  L.~F.~Abbott and P.~Sikivie,
  Phys.\ Lett.\ B {\bf 120} (1983) 133.
\bibitem{Dine:1982ah}
  M.~Dine and W.~Fischler,
  Phys.\ Lett.\ B {\bf 120} (1983) 137.
  
\bibitem{Vicari:2008jw} 
  E.~Vicari and H.~Panagopoulos,
  Phys.\ Rept.\  {\bf 470}, 93 (2009)
  [arXiv:0803.1593 [hep-th]].


\bibitem{Gross:1980br} 
  D.~J.~Gross, R.~D.~Pisarski and L.~G.~Yaffe,
  Rev.\ Mod.\ Phys.\  {\bf 53}, 43 (1981).
\bibitem{Schafer:1996wv} 
  T.~Sch\"{a}fer and E.~V.~Shuryak,
  Rev.\ Mod.\ Phys.\  {\bf 70}, 323 (1998)
  [hep-ph/9610451].
\bibitem{Morris:1984zi}  
  T.~R.~Morris, D.~A.~Ross and C.~T.~Sachrajda,
  Nucl.\ Phys.\ B {\bf 255}, 115 (1985).
\bibitem{Ringwald:1999ze} 
  A.~Ringwald and F.~Schrempp,
  Phys.\ Lett.\ B {\bf 459}, 249 (1999)
  [hep-lat/9903039].
\bibitem{Borsanyi:2015cka} 
  S.~Borsanyi {\it et al.},
  Phys.\ Lett.\ B {\bf 752}, 175 (2016)
  [arXiv:1508.06917 [hep-lat]].


\bibitem{Bonati:2013tt} 
  C.~Bonati, M.~D'Elia, H.~Panagopoulos and E.~Vicari,
  Phys.\ Rev.\ Lett.\  {\bf 110}, 252003 (2013)
  [arXiv:1301.7640 [hep-lat]].
\bibitem{Bonati:2015uga} 
  C.~Bonati,
  JHEP {\bf 1503}, 006 (2015)
  [arXiv:1501.01172 [hep-lat]].
\bibitem{Xiong:2015dya} 
  G.~Y.~Xiong, J.~B.~Zhang, Y.~Chen, C.~Liu, Y.~B.~Liu and J.~P.~Ma,
  Phys.\ Lett.\ B {\bf 752}, 34 (2016)
  [arXiv:1508.07704 [hep-lat]].



\bibitem{Alles:1996nm} 
  B.~Alles, M.~D'Elia and A.~Di Giacomo,
  Nucl.\ Phys.\ B {\bf 494}, 281 (1997)
  [Erratum Nucl.\ Phys.\ B {\bf 679}, 397 (2004)]
  [hep-lat/9605013].
\bibitem{Alles:2000cg} 
  B.~Alles, M.~D'Elia and A.~Di Giacomo,
  Phys.\ Lett.\ B {\bf 483}, 139 (2000)
  [hep-lat/0004020].
\bibitem{Gattringer:2002mr} 
  C.~Gattringer, R.~Hoffmann and S.~Schaefer,
  Phys.\ Lett.\ B {\bf 535}, 358 (2002)
  [hep-lat/0203013].
\bibitem{Lucini:2004yh} 
  B.~Lucini, M.~Teper and U.~Wenger,
  Nucl.\ Phys.\ B {\bf 715}, 461 (2005)
  [hep-lat/0401028].
\bibitem{DelDebbio:2004rw} 
  L.~Del Debbio, H.~Panagopoulos and E.~Vicari,
  JHEP {\bf 0409}, 028 (2004)
  [hep-th/0407068].
\bibitem{Berkowitz:2015aua} 
  E.~Berkowitz, M.~I.~Buchoff and E.~Rinaldi,
  Phys.\ Rev.\ D {\bf 92}, 034507 (2015)
  [arXiv:1505.07455 [hep-ph]].


\bibitem{Bazavov:2010xr} 
  A.~Bazavov {\it et al.} [MILC Collaboration],
  Phys.\ Rev.\ D {\bf 81}, 114501 (2010)
  [arXiv:1003.5695 [hep-lat]].
\bibitem{Bazavov:2012xda} 
  A.~Bazavov {\it et al.} [MILC Collaboration],
  Phys.\ Rev.\ D {\bf 87}, 054505 (2013)
  [arXiv:1212.4768 [hep-lat]].
\bibitem{Cichy:2013rra} 
  K.~Cichy {\it et al.} [ETM Collaboration],
  JHEP {\bf 1402}, 119 (2014)
  [arXiv:1312.5161 [hep-lat]].
\bibitem{Bruno:2014ova} 
  M.~Bruno, S. Schaefer and R. Sommer [ALPHA Collaboration],
  JHEP {\bf 1408}, 150 (2014)
  [arXiv:1406.5363 [hep-lat]].
\bibitem{Fukaya:2014zda} 
  H.~Fukaya {\it et al.} [JLQCD Collaboration],
  PoS LATTICE {\bf 2014}, 323 (2014)
  [arXiv:1411.1473 [hep-lat]].

\bibitem{Alles:1996vn} 
  B.~Alles, G.~Boyd, M.~D'Elia, A.~Di Giacomo and E.~Vicari,
  Phys.\ Lett.\ B {\bf 389}, 107 (1996)
  [hep-lat/9607049].
\bibitem{DelDebbio:2002xa} 
  L.~Del Debbio, H.~Panagopoulos and E.~Vicari,
  JHEP {\bf 0208}, 044 (2002)
  [hep-th/0204125].
\bibitem{DelDebbio:2004xh} 
  L.~Del Debbio, G.~M.~Manca and E.~Vicari,
  Phys.\ Lett.\ B {\bf 594}, 315 (2004)
  [hep-lat/0403001].
\bibitem{Schaefer:2010hu} 
  S.~Schaefer {\it et al.} [ALPHA Collaboration],
  Nucl.\ Phys.\ B {\bf 845}, 93 (2011)
  [arXiv:1009.5228 [hep-lat]].
\bibitem{Kitano:2015fla} 
  R.~Kitano and N.~Yamada,
  JHEP {\bf 1510}, 136 (2015)
  [arXiv:1506.00370 [hep-ph]].

\bibitem{weisz} 
  P.~Weisz,
  Nucl.\ Phys.\ B {\bf 212}, 1 (1983).
\bibitem{curci} 
G.~Curci, P.~Menotti and G.~Paffuti,
  Phys.\ Lett.\ B {\bf 130}, 205 (1983)
  [Erratum-ibid.\ B {\bf 135}, 516 (1984)].

\bibitem{Morningstar:2003gk} 
  C.~Morningstar and M.~J.~Peardon,
  Phys.\ Rev.\ D {\bf 69}, 054501 (2004)
  [hep-lat/0311018].

\bibitem{physline1}
  Y.~Aoki, S.~Borsanyi, S.~Durr, Z.~Fodor, S.~D.~Katz, S.~Krieg and K.~K.~Szabo,
  JHEP {\bf 0906}, 088 (2009)
  [arXiv:0903.4155 [hep-lat]]. 
\bibitem{physline2}
  S.~Borsanyi, G.~Endrodi, Z.~Fodor, A.~Jakovac, S.~D.~Katz, S.~Krieg, C.~Ratti and K.~K.~Szabo,
  JHEP {\bf 1011}, 077 (2010)
  [arXiv:1007.2580 [hep-lat]]; 
\bibitem{physline3}
  S.~Borsanyi, Z.~Fodor, C.~Hoelbling, S.~D.~Katz, S.~Krieg and K.~K.~Szabo,
  Phys.\ Lett.\ B {\bf 730}, 99 (2014)
  [arXiv:1309.5258 [hep-lat]].

\bibitem{Neuberger:1997fp} 
  H.~Neuberger,
  Phys.\ Lett.\ B {\bf 417}, 141 (1998)
  [hep-lat/9707022].
\bibitem{Hasenfratz:1998ri} 
  P.~Hasenfratz, V.~Laliena and F.~Niedermayer,
  Phys.\ Lett.\ B {\bf 427}, 125 (1998)
  [hep-lat/9801021].
\bibitem{Luscher:1998pqa} 
  M.~Luscher,
  Phys.\ Lett.\ B {\bf 428}, 342 (1998)
  [hep-lat/9802011].
\bibitem{Luscher:2004fu} 
  M.~Luscher,
  Phys.\ Lett.\ B {\bf 593}, 296 (2004)
  [hep-th/0404034].

\bibitem{Giusti:2008vb} 
  L.~Giusti and M.~Luscher,
  JHEP {\bf 0903}, 013 (2009)
  [arXiv:0812.3638 [hep-lat]].

\bibitem{cooling}
  B.~Berg,
  Phys.\ Lett.\ B {\bf 104}, 475 (1981);
  Y.~Iwasaki and T.~Yoshie,
  Phys.\ Lett.\ B {\bf 131}, 159 (1983);
  S.~Itoh, Y.~Iwasaki and T.~Yoshie,
  Phys.\ Lett.\ B {\bf 147}, 141 (1984);
  M.~Teper,
  Phys.\ Lett.\ B {\bf 162}, 357 (1985);
  E.~-M.~Ilgenfritz {\it et al.},
  Nucl.\ Phys.\ B {\bf 268}, 693 (1986).

\bibitem{Luscher:2009eq} 
  M.~Luscher,
  Commun.\ Math.\ Phys.\  {\bf 293}, 899 (2010)
  [arXiv:0907.5491 [hep-lat]].
\bibitem{Luscher:2010iy} 
  M.~Luscher,
  JHEP {\bf 1008}, 071 (2010)
  [JHEP {\bf 1403}, 092 (2014)]
  [arXiv:1006.4518 [hep-lat]].


\bibitem{Bonati:2014tqa} 
  C.~Bonati and M.~D'Elia,
  Phys.\ Rev.\ D {\bf 89}, 105005 (2014)
  [arXiv:1401.2441 [hep-lat]].
\bibitem{Cichy:2014qta} 
  K.~Cichy, A.~Dromard, E.~Garcia-Ramos, K.~Ottnad, C.~Urbach, M.~Wagner, U.~Wenger and F.~Zimmermann,
  PoS LATTICE {\bf 2014}, 075 (2014)
  [arXiv:1411.1205 [hep-lat]].
\bibitem{Namekawa:2015wua} 
  Y.~Namekawa,
  PoS LATTICE {\bf 2014}, 344 (2015)
  [arXiv:1501.06295 [hep-lat]].
\bibitem{Alexandrou:2015yba} 
  C.~Alexandrou, A.~Athenodorou and K.~Jansen,
  Phys.\ Rev.\ D {\bf 92}, 125014 (2015)
  [arXiv:1509.04259 [hep-lat]].

\bibitem{Sommer:2014mea} 
  R.~Sommer,
  PoS LATTICE {\bf 2013}, 015 (2014)
  [arXiv:1401.3270 [hep-lat]].
\bibitem{Fodor:2014cpa} 
  Z.~Fodor, K.~Holland, J.~Kuti, S.~Mondal, D.~Nogradi and C.~H.~Wong,
  JHEP {\bf 1409}, 018 (2014)
  [arXiv:1406.0827 [hep-lat]].


\bibitem{DiVecchia:1981qi} 
  P.~Di Vecchia, K.~Fabricius, G.~C.~Rossi and G.~Veneziano,
  Nucl.\ Phys.\ B {\bf 192}, 392 (1981).

\bibitem{Bonati:2015sqt} 
  C.~Bonati, M.~D'Elia and A.~Scapellato,
  Phys.\ Rev.\ D {\bf 93}, 025028 (2016)
  [arXiv:1512.01544 [hep-lat]].


\bibitem{DiVecchia:1980ve}
  P.~Di Vecchia and G.~Veneziano,
  Nucl.\ Phys.\ B {\bf 171} (1980) 253.
\bibitem{Leutwyler:1992yt} 
  H.~Leutwyler and A.~V.~Smilga,
  Phys.\ Rev.\ D {\bf 46}, 5607 (1992).
\bibitem{Mao:2009sy}
  Y.~Y.~Mao {\it et al.} [TWQCD Collaboration],
  Phys.\ Rev.\ D {\bf 80} (2009) 034502
  [arXiv:0903.2146 [hep-lat]].
\bibitem{Guo:2015oxa}
  F.~K.~Guo and U.~G.~Meißner,
  Phys.\ Lett.\ B {\bf 749} (2015) 278
  [arXiv:1506.05487 [hep-ph]].
\bibitem{diCortona:2015ldu}
  G.~G.~di Cortona, E.~Hardy, J.~P.~Vega and G.~Villadoro,
  JHEP {\bf 1601}, 034 (2016)
  [arXiv:1511.02867 [hep-ph]].


\bibitem{DelDebbio:2004ns} 
  L.~Del Debbio, L.~Giusti and C.~Pica,
  Phys.\ Rev.\ Lett.\  {\bf 94}, 032003 (2005)
  [hep-th/0407052].
\bibitem{Durr:2006ky} 
  S.~Durr, Z.~Fodor, C.~Hoelbling and T.~Kurth,
  JHEP {\bf 0704}, 055 (2007)
  [hep-lat/0612021].
\bibitem{Luscher:2010ik} 
  M.~Luscher and F.~Palombi,
  JHEP {\bf 1009}, 110 (2010)
  [arXiv:1008.0732 [hep-lat]].
\bibitem{Cichy:2015jra} 
  K.~Cichy {\it et al.} [ETM Collaboration],
  JHEP {\bf 1509}, 020 (2015)
  [arXiv:1504.07954 [hep-lat]].
\bibitem{Ce:2015qha} 
  M.~C\'e, C.~Consonni, G.~P.~Engel and L.~Giusti,
  Phys.\ Rev.\ D {\bf 92}, 074502 (2015)
  [arXiv:1506.06052 [hep-lat]].



\bibitem{D'Elia:2003gr} 
  M.~D'Elia,
  Nucl.\ Phys.\ B {\bf 661}, 139 (2003)
  [hep-lat/0302007].
\bibitem{Giusti:2007tu} 
  L.~Giusti, S.~Petrarca and B.~Taglienti,
  Phys.\ Rev.\ D {\bf 76}, 094510 (2007)
  [arXiv:0705.2352 [hep-th]].
\bibitem{Panagopoulos:2011rb} 
  H.~Panagopoulos and E.~Vicari,
  JHEP {\bf 1111}, 119 (2011)
  [arXiv:1109.6815 [hep-lat]].

\bibitem{Buchoff:2013nra} 
  M.~I.~Buchoff {\it et al.},
  Phys.\ Rev.\ D {\bf 89}, 054514 (2014)
  [arXiv:1309.4149 [hep-lat]].

\bibitem{Trunin:2015yda} 
  A.~Trunin, F.~Burger, E.-M.~Ilgenfritz, M.~P.~Lombardo and M.~M\"uller-Preussker,
  arXiv:1510.02265 [hep-lat].

\bibitem{Wantz:2009it}
  O.~Wantz and E.~P.~S.~Shellard,
  Phys.\ Rev.\ D {\bf 82} (2010) 123508
  [arXiv:0910.1066 [astro-ph.CO]].

\bibitem{Ade:2015xua}
  P.~A.~R.~Ade {\it et al.} [Planck Collaboration],
  arXiv:1502.01589 [astro-ph.CO].

\bibitem{Callan:1977gz} 
  C.~G.~Callan, Jr., R.~F.~Dashen and D.~J.~Gross,
  Phys.\ Rev.\ D {\bf 17}, 2717 (1978).

\bibitem{deForcrand:1997fm} 
  P.~de Forcrand, J.~E.~Hetrick, T.~Takaishi and A.~J.~van der Sijs,
  Nucl.\ Phys.\ Proc.\ Suppl.\  {\bf 63}, 679 (1998)
  [hep-lat/9709104].
\bibitem{openbc} 
  M.~Luscher and S.~Schaefer,
  JHEP {\bf 1107}, 036 (2011)
  [arXiv:1105.4749 [hep-lat]].
\bibitem{surfing} 
  A.~Laio, G.~Martinelli and F.~Sanfilippo,
  arXiv:1508.07270 [hep-lat].
\bibitem{endres}
  M.~G.~Endres, R.~C.~Brower, W.~Detmold, K.~Orginos and A.~V.~Pochinsky,
  Phys.\ Rev.\ D {\bf 92}, 114516 (2015)
  [arXiv:1510.04675 [hep-lat]].

\end{thebibliography}
\end{document}